\documentclass[aps,prb,twocolumn]{revtex4}

\usepackage[dvips]{graphicx}

\begin{document}

\title{Schottky barriers at metal-finite semiconducting carbon nanotube 
interfaces}



\author{ Yongqiang Xue $^{*}$ and Mark A. Ratner}
\affiliation{Department of Chemistry and Materials Research Center, 
Northwestern University, Evanston, IL 60208}
\date{\today}

\begin{abstract}
Electronic properties of metal-finite semiconducting carbon nanotube 
interfaces are studied as a function of the nanotube length using a 
self-consistent tight-binding theory. We find that the shape of the 
potential barrier depends on the long-range tail of the charge transfer, 
leading to an injection barrier thickness comparable to half of the 
nanotube length until the nanotube reaches the bulk limit. The conductance 
of the nanotube junction shows a transition from tunneling to 
thermally-activated transport with increasing nanotube length.  
\end{abstract}


\maketitle


Single-wall carbon nanotubes (SWNT) are ideal systems for 
studying transport in the length scale ranging from the molecular limit as 
all-carbon cylindrical molecules to the bulk limit as quasi-one-dimensional 
conductors,~\cite{Dekker99} and many device concepts have been sucessfully 
demonstrated on a single-tube basis.\cite{Dekker,Lieber00,Avouris02}
Among the device physics problems raised,  the nature of the 
electronic transport through a metal-semiconducting carbon nanotube 
interface stands out ~\cite{Xue99,TersoffNT,Odin00,De02} 
as one of the basic device building blocks.~\cite{Rhod} 
Although transport through a metal-long carbon nanotube interface has been 
studied using the bulk band structure and one-dimensional 
electrostatics,~\cite{TersoffNT,Odin00,De02} it 
is important to investigate device functionality of finite carbon nanotubes 
with lengths of nanometer range and three-dimensional electrostatics, which  
will also shed light on the scaling limit of carbon nanotube devices. 
 
\begin{figure}
\vspace{3.0cm} 
\includegraphics[height=3.0in,width=3.0in]{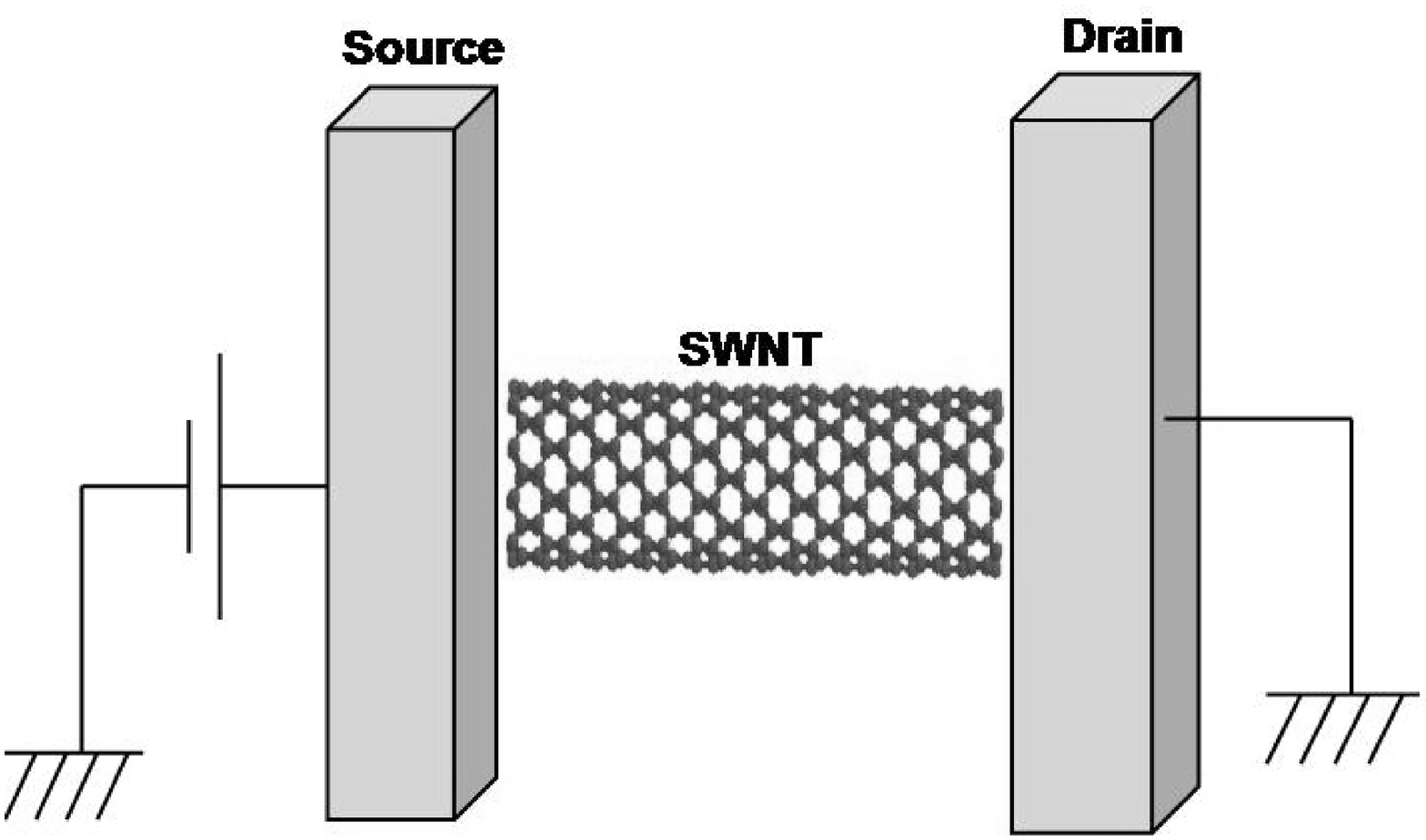}
\caption{\label{xueFig1} Schematic illustration of the metal-SWNT-metal junction. }
\vspace{2.0cm} 
\end{figure}

Here, we analyze the evolution of the electronic properties of the 
metal-finite SWNT interface as the length is varied from the molecular 
to the bulk limit. The model system ia illustrated schematically in 
Fig.\ \ref{xueFig1}, where the open-ended 
SWNT molecule is attached to the surfaces of the metallic electrodes 
through the dangling bonds at the end. 
The device structure chosen represents 
an atomic-scale analogue of the metal-semiconductor interface since both 
the interface area and the active device region are atomic-scale.   
We find that  the shape of the potential barrier depends on the charge 
transfer throughout the junction, which leads to an injection barrier 
thickness comparable to half of the nanotube length until the nanotube 
reaches the bulk limit. As a consequence, the SWNT junction conductance 
shows a transition from tunneling to thermally-activated transport as the 
nanotube length increases.  

We take $(10,0)$ SWNT as the protype semiconducting SWNT,  
whose work function is taken as that of the graphite ($4.5$ eV).~\cite{Dekker} 
We consider gold (Au) and titanium (Ti) electrodes as examples of high- 
and low- work function metals ($5.1$ and $4.33$ eV respectively for 
polycrystalline materials). We describe the electronic structure of the isolated 
SWNT using the Extended Huckel Theory (EHT) with non-orthogonal basis 
sets $\phi_{m}(\vec r)$, ~\cite{Hoffmann88} which gives a 
bulk (infinitely long) band gap of $\approx 0.9(eV)$ for $(10,0)$ SWNT. 
Since the SWNT Fermi level is located approximately at mid-gap, 
before the contact formation, the gold Fermi-level ($-5.1$ eV) lies 
below the valence band edge while the titanium Fermi-level ($-4.33$ eV) 
lies between the mid-gap and conduction band edge of the bulk SWNT.  

Since the screening of Coulomb interaction is relatively ineffective within the 
SWNT due to the reduced dimensionality,~\cite{TersoffNT,Odin00} an 
atomistic study of the electronic processes throughout the 
metal-SWNT-metal junction is needed. We use a self-consistent 
tight-binding theory based on the semi-empirical implementation of 
the self-consistent Matrix Green's function (SCMGF) method for 
first-principles modeling of molecular-scale devices.~\cite{Xue02} 
Given the EHT Hamiltonian $H_{0}$ of the isolated SWNT,  we 
calculate the density matrix $\rho_{ij}$ and therefore the electron density 
of the equilibrium SWNT junction from
\begin{eqnarray}
\label{GE}
G^{R} 
&=& \{ (E+i0^{+})S-H-\Sigma_{L}(E)-\Sigma_{R}(E) \}^{-1}, \\
\rho &=& \int \frac{dE}{2\pi }Imag[G^{R}](E)f(E-E_{F}).
\end{eqnarray}
where the effect of coupling to the electrodes are included as matrix 
self-energy operators $\Sigma_{L(R)}$ and calculated using tight-binding 
parameters (we use a nanotube end-surface 
distance of $ 2.5 \AA $ here).~\cite{Xue02,Papa86} Here $S$ is overlap matrix and 
$f(E-E_{F})$ is the Fermi distribution describing the electrodes 
(see Ref.\ \onlinecite{Xue02} for details). The SWNT Hamiltonian 
is now $H=H_{0}+\delta V[\delta \rho]$ where $\delta \rho$ is the 
density of transferred charge and $\delta V$ 
is the induced change in the electrostatic potential. 

To proceed with self-consistent calculation,  we approximate the charge 
distribution as superposition of atom-centered charge distributions~\cite{Frau98} 
$\delta \rho(\vec r)=\sum_{i} \delta N_{i} \rho_{i}(\vec r-\vec r_{i})$,  
where $\delta N_{i}=(\rho S)_{ii}-N_{i}^{0}$ and $N_{i}^{0}$ is the number of 
valence electrons on atomic-site $i$ of the bare SWNT. 
$\rho_{i}(\vec r)=\frac{1}{N_{\zeta_{i}}} e^{-\zeta_{i} r}$ is a normalized 
Slater-type function ($\int d\vec r \rho_{i}(\vec r)=1$).~\cite{Frau98} The 
exponent $\zeta_{i}$ is chosen such that 
$\int d\vec r d\vec r' \rho_{i}(\vec r) \rho_{i}(\vec r')/|\vec r- \vec r'|  
=I_{i}-A_{i}$~\cite{Frau98}, where $I_{i}(A_{i})$ are the atomic electron 
affinity (ionization potential). In this way, we obtain $\delta V(\vec r)
=\sum_{i} \delta N_{i}V_{i}(\vec r -\vec r_{i})$, where 
$V_{i}=\int d\vec r' \rho_{i}(\vec r'-\vec r_{i})/|\vec r- \vec r'|$ 
can be evaluated analytically~\cite{Frau98}. 
We take into account the image-potential effect by including within 
$\delta \rho$ both the tranfered-charge on the carbon atoms and
their image charges, rather than imposing an image-type potential 
correction. The matrix elements of the potential 
$\delta V_{mn}=\int d\vec r \phi_{m}^{*}(\vec r)\delta V(\vec r)\phi_{n}(\vec r)$  
are calculated using two types of scheme: (1) If $m,n$ belong to 
the same atomic site $i$, we calculate it by direct numerical integration; 
(2) if $m,n$ belong to different atomic sites, we use the 
approximation $\delta V_{mn}=S_{mn}(\delta V_{mm}+\delta V_{nn})/2$. 

\begin{figure}
\includegraphics[height=3.0in,width=3.0in]{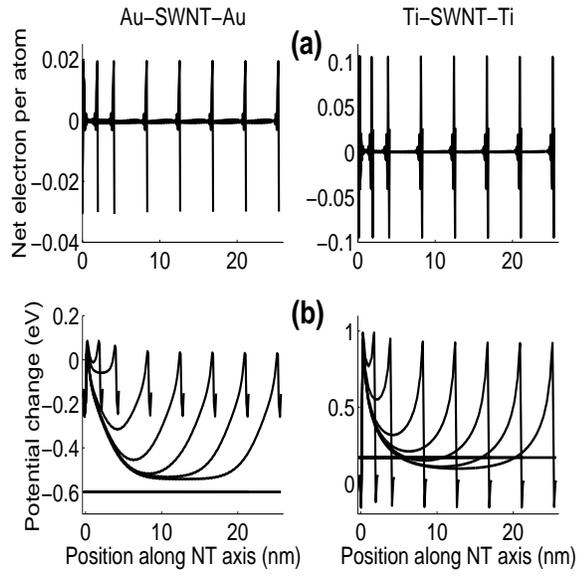}
\caption{\label{xueFig2}  Charge transfer (a) and electrostatic potential change 
(b) at the metal-finite SWNT-metal junction as a function of SWNT length 
for seven different lengths. The horizontal lines in (b) denote the 
work function differences between the electrodes and the bulk SWNT.  }  
\end{figure}

\begin{figure}
\includegraphics[height=3.0in,width=3.0in]{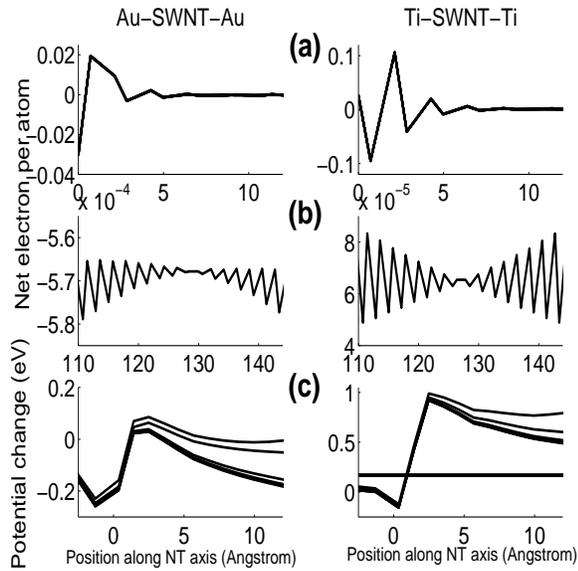}
\caption{\label{xueFig3}  Magnified view of transferred-charge at the 
metal-SWNT interface (a) 
and in the middle of the 60-unitcell SWNT (b). The magnitude of the 
interfacial charge transfer is approximately identical for all finite SWNTs studied.  
(c) shows magnified view of the potential shift at the interface.  }
\end{figure}

The finite SWNT lengths investigated are $2.0,4.1,
8.4,12.6,16.9,21.2$ and $25.4$ (nm), which corresponds to number of 
 unitcells of $5,10,20,30,40,50$ and $60$ respectively and spans the 
entire range from the molecular limit to the bulk limit.
Due to the dangling bonds at the end, charge transfer occurs 
between the end and the interior carbon atoms of 
the SWNT, which should be corrected self-consistently first and give the initial 
charge configuration $N^{0}_{i}$. ~\cite{Xue02} The calculated 
charge transfer and electrostatic potential change for the Au-SWNT-Au 
and Ti-SWNT-Ti junctions are shown in Fig. \ref{xueFig2}.  
The magnified view of the transferred charge and potential 
shift at the interface and in the middle of the SWNT appear 
in Fig.\ \ref{xueFig3}. 

The charge transfer at the metal-SWNT interface reflects the 
bonding configuration change upon contact to the metallic surfaces, 
involving mainly the end carbon atoms and decaying rapidly into 
the interior of the SWNT molecule. ~\cite{Xue02} 
The electrostatic potential change is instead determined by the transferred 
charge throughout the metal-SWNT-junction due to the long-range 
Coulomb interaction, so its magnitude in the middle of the SWNT 
increases with increasing length until the finite SWNT reaches the bulk limit 
despite the small magnitude of the transferred charge in the middle of the 
SWNT.~\cite{Note} This gives a maximal barrier thickness for 
electron injection of roughly $10(nm)$, i.e., half of the nanotube 
length where it reaches the bulk limit ($50$ unitcells). The magnitude of 
both the charge transfer 
and the potential shift at the metal-SWNT interface are approximately 
the same for all the finite SWNTs studied. 

For the SWNT molecules which have approached the bulk limit, the 
interface perturbation of the electron states in the middle can be essentially 
neglected. Charging and ``band'' shift is determined by the shift of the 
local density of states (LDOS) in the middle of the SWNT relative to 
the metal Fermi-level. Note that due to the 
three-dimensional electrostatics of the metal-SWNT-metal junction, the 
change in the electrostatic potential energy induced by the transferred charge 
decays moving away from the cylindrical surface of the SWNT. 
Therefore \emph{the shift of the LDOS along the SWNT axis doesn't follow 
the shift in the potential energy}, different from the bulk 
metal-semiconductor interface. For the 60-unitcell SWNT, we find that 
for the Au-SWNT-Au junction 
the Fermi-level is located slightly below (by $\sim 0.05 $eV) the mid-gap, 
while for the Ti-SWNT-Ti junction it is located above (by $\sim 0.15 $eV) 
the mid-gap. This is consistent with the fact that in the interior of the 
SWNT the Fermi-level must lie within the band gap to ensure the small 
perturbation of the electron states there.  

These charge transfer processes are often characterized 
as ``charge-transfer doping''. For the Au-SWNT-Au (Ti-SWNT-Ti) junction, 
there is a small positive (negative) charge transfer 
of $5.7 \times 10^{-4}$ ($-6.4 \times 10^{-5}$) per atom in the 
middle of the $60$-unitcell SWNT. The oscillation of transferred-charge in the 
middle of the SWNT is due to the two-sublattice structure of zigzag tube, leading  
to charge transfer among atoms within the unitcell. The SWNT is therefore 
``hole-doped'' by contacting to high work function (Au) and ``electron-doped'' 
by contacting to low work function (Ti) electrode. However, 
within the coherent transport regime, it is clear that both the charge transfer 
at the interface and the charge transfer inside the SWNT contribute only 
indirectly to electron transport  by modulating the potential landscape acrosss 
the metal-SWNT-metal junction.  

\begin{figure}
\includegraphics[height=3.0in,width=3.0in]{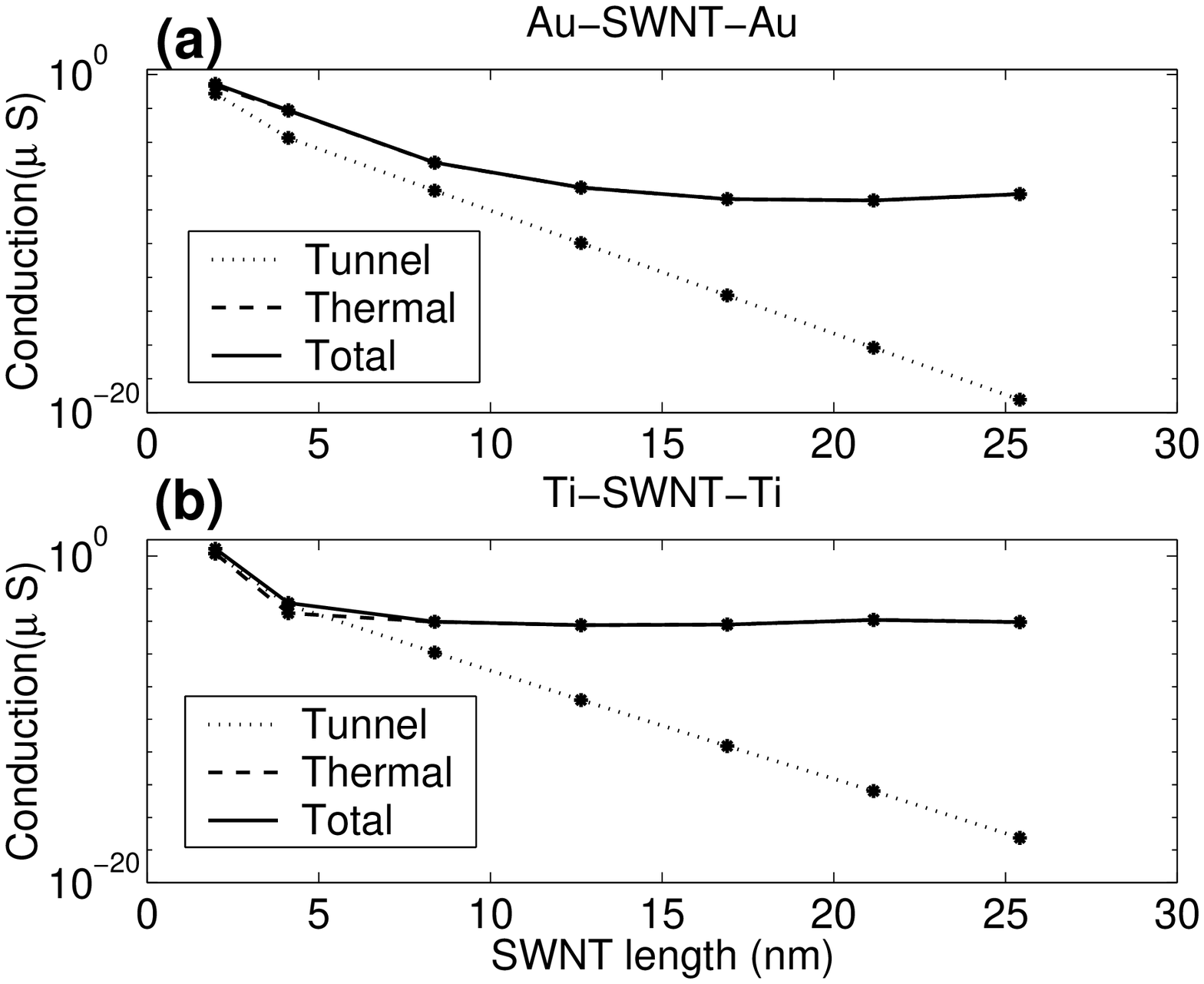}
\caption{\label{xueFig4}  Room temperature conductance of the 
metal-finite SWNT-metal junction as a function of SWNT length. }
\end{figure}

Given the potential shift across the metal-SWNT interface, we 
can evaluate the length and temperature dependence of the SWNT junction 
conductance using the Landauer formula
\begin{equation}
\label{GT}
G=\frac{2e^{2}}{h}\int dE T(E)[-\frac{df}{dE}(E-E_{F})]=G_{Tu}+G_{Th}
\end{equation} 
where the transmission is calculated from 
$T(E)=Tr[\Gamma_{L}(E)G^{R}(E)\Gamma_{R}(E)G^{A}(E)]$.~\cite{Xue02}  
Here we have separated the conductance  into tunneling contribution 
$G_{Tu}=\frac{2e^{2}}{h}T(E_{F})$ and thermal-activation 
contribution $G_{Th}=G-G_{Tu}$. 
The result for room temperature conductance is shown in Fig.\  \ref{xueFig4}. 
The tunneling conductance (also the zero temperature conductance) 
for both junctions decreases exponentially with the SWNT length for 
SWNT longer than $4.1(nm)$ appropriate for tunneling transport 
though potential barriers with identical barrier height. The room-temperature 
conductance instead saturates with increasing SWNT length. This is because 
the tunneling is exponentially suppressed for the longer SWNT molecules, while 
the transport becomes dominated by thermal-acitivation over the top of the 
potential barrier, whose height is approximately the same for all the finite SWNTs 
investigated.  For Ti-SWNT-Ti junction, this leads to a transition from 
tunneling to thermally-activated transport at roughly $5(nm)$. For Au-SWNT-Au 
junction, the thermal contribution is already larger than the tunneling contribution 
for the smallest SWNT studied at room temperature. 

The long-range nature of the charge transfer and potential barrier suggests the 
gate-modulation of junction conductance in SWNT-based transistors may be 
achieved through the modulation of the injection barrier at the metal-natotube 
interface, in agreement with recent experimental works.~\cite{Avouris02} 
The SWNT junction transport characteristics are sensitive to the shape of the 
potential change induced by both the gate voltage and source-drain voltage, 
whose effect increases with increasing SWNT length. Further analysis are 
needed to achieve a through understanding of nanotube-based devices.   

This work was supported by the DARPA Molectronics program, 
the DoD-DURINT program and the NSF Nanotechnology Initiative. 


%




  

\newpage

\vspace{3.0cm} 

\begin{references}
\bibitem[*]{Xue} Corresponding author. Email: ayxue@chem.nwu.edu 
\bibitem{Dekker99} C. Dekker, Phys.\ Today {\bf 52}(5), 22 (1999).
\bibitem{Dekker} S.J. Tans, A.R.M. Vershueren and C. Dekker, 
Nature {\bf 393}, 49 (1998); A. Bachtold, P. Hadley, T. Nakanishi and 
C. Dekker, Science {\bf 294}, 1317 (2001).
\bibitem{Lieber00} T. Rueckes, K. Kim, E. Joslevich, G. Tseng, C. Cheung 
and C.M. Lieber, Science {\bf 289}, 94 (2000).
\bibitem{Avouris02} V. Derycke, R. Martel, J. Appenzeller and Ph. Avouris, 
Appl.\ Phys.\ Lett.\ {\bf 80}, 2773 (2002); S. Heinze, J. Tersoff, R. Martel, 
V. Derycke, J. Appenzeller and Ph. Avouris, 
Phys.\ Rev.\ Lett.\ {\bf 89}, 106801 (2002).
\bibitem{Xue99} Y. Xue and S. Datta, Phys.\ Rev.\ Lett.\ {83}, 4844 (1999).
\bibitem{TersoffNT} F. Le{\'o}nard and J. Tersoff, Phys.\ Rev.\ Lett.\ 
{\bf 84}, 4693 (2000); Appl.\ Phys.\ Lett.\ {\bf 81}, 4835 (2002). 
\bibitem{Odin00} A.A. Odinsov, Phys.\ Rev.\ Lett.\ {85}, 150 (2000).  
\bibitem{De02} T. Nakanishi, A. Bachtold and C. Dekker, 
Phys.\ Rev.\ B {\bf 66}, 73307 (2002).
\bibitem{Rhod} E.H. Rhoderick and R.H. Williams, 
\emph{Metal-Semiconductor Contacts}, 2nd edition 
(Clarendon Press, Oxford, 1988).
\bibitem{Xue02} Y. Xue, S. Datta and M. A. Ratner, Chem.\ Phys.\ {\bf 281}, 
151 (2002); Y. Xue and M.A. Ratner, MRS Proceedings {\bf 734}, B6.8 (2003); 
cond-mat/0303179.  
\bibitem{Papa86} D. A. Papaconstantopoulos, \emph{Handbook of the Band 
Structure of Elemental Solids} (Plenum Press, New York, 1986).
\bibitem{Frau98} M. Elstner, D. Porezag, G. Jungnickel, J. Elsner, M. Haugk, 
Th. Frauenheim, S. Suhai and G. Seifert, Phys.\ Rev.\ B {\bf 58}, 7260 (1998).
\bibitem{Hoffmann88} R. Hoffmann, Rev.\ Mod.\ Phys.\ {\bf 60}, 601 (1988); 
A. Rochefort, D.R. Salahub and Ph. Avouris, 
J.\ Phys.\ Chem.\ B {\bf 103}, 641 (1999). 
\bibitem{Note} A quantum capacitance of $\sim 0.3$ pF/cm for the SWNT 
can be estimated from the differences in the potential change and 
transferred charge in the middle of the $50$- and $60$-unitcell SWNTs. 
\end{references}
\end{document}